\documentclass[8.5pt,twoside,twocolumn]{article}
\usepackage{ulem} 
\usepackage[super,sort&compress,comma]{natbib} 
\usepackage{mhchem}
\usepackage{times,mathptmx}
\usepackage{sectsty}
\usepackage{balance} 
\usepackage{graphicx}
\usepackage{lastpage}
\usepackage[format=plain,justification=raggedright,singlelinecheck=false,font=small,labelfont=bf,labelsep=space]{caption} 
\usepackage{fancyhdr}
\usepackage{url}
\usepackage{color}
\usepackage{wasysym}

\begin{document}
\fancyhead{}
\renewcommand{\headrulewidth}{1pt} 
\renewcommand{\footrulewidth}{1pt}
\setlength{\arrayrulewidth}{1pt}
\setlength{\columnsep}{6.5mm}
\setlength\bibsep{1pt}
\twocolumn[
\begin{@twocolumnfalse}
\noindent\LARGE{\textbf{Capillary levelling of a cylindrical hole in a viscous film}}
\vspace{0.6cm}

\noindent\large{\textbf{Matilda Backholm,\textit{$^{a}$} Michael Benzaquen,\textit{$^{b}$} Thomas Salez,\textit{$^{b}$} Elie Rapha\"el,\textit{$^{b}$} and Kari Dalnoki-Veress$^{\ast}$\textit{$^{a,b}$}}}\vspace{0.5cm}

\noindent\textit{\small{\textbf{Received Xth XXXXXXXXXX 20XX, Accepted Xth XXXXXXXXX 20XX\newline
First published on the web Xth XXXXXXXXXX 200X}}}

\noindent \textbf{\small{DOI: 10.1039/b000000x}}
\vspace{0.6cm}

\noindent \normalsize{The capillary levelling of cylindrical holes in viscous polystyrene films was studied using atomic force microscopy as well as quantitative analytical scaling arguments based on thin film theory and self-similarity. The relaxation of the holes was shown to consist of two different time regimes: an early regime where opposing sides of the hole do not interact, and a late regime where the hole is filling up. For the latter, the self-similar asymptotic profile was derived analytically and shown to be in excellent agreement with experimental data. Finally, a binary system of two holes in close proximity was investigated where the individual holes fill up at early times and coalesce at longer times.}
\vspace{0.5cm}
\end{@twocolumnfalse}]
 
\section*{Introduction}
\footnotetext{\textit{$^{a}$~Department of Physics \& Astronomy and the Brockhouse Institute for Materials Research, McMaster University, Hamilton, Canada. Fax: 905 521 2773; Tel: 905 525 9140 ext. 22658; E-mail: dalnoki@mcmaster.ca}}
\footnotetext{\textit{$^{b}$~Laboratoire de Physico-Chimie Th\'eorique, UMR CNRS 7083 Gulliver, ESPCI, Paris, France. }}

The nanorheology of thin polymer films has recently received significant scientific attention due to the important effect confinement has on the mobility of the polymer molecules\cite{Gra94, For96, Si05, Bod06, Shi07, Fak08, Bae09, Rae10, Tho11, Li12}. The understanding of the viscous flow and stability of these films is important for technological applications, where ultra-thin polymer coatings are frequently used in, e.g., nanolithography\cite{Tei11} and nanoimprinting\cite{Cho95}, as well as in the development of non-volatile computer memory devices\cite{Ouy04}.
\bigskip

This topic is also of importance for applied mathematics and fundamental physics. The use of dimensional analysis\cite{Buckingham1914}, scaling, and self-similarity has led to remarkable historical results in hydrodynamics\cite{Reynolds1895} and particularly in the study of turbulence\cite{Kolmogorov1942,Kolmogorov1941}. An interesting example is that of the prediction of the nuclear explosion power by Taylor\cite{Taylor1950a, Taylor1950b}. The theory of intermediate asymptotics developed by Barenblatt\cite{Barenblatt1996} investigates and makes use of the more profound meaning of these concepts to provide a general method for solving nonlinear problems at intermediate times. The term ``intermediate" refers to time scales that are large enough for the system to have forgotten the initial conditions but also far enough from the generally predictable final equilibrium steady state. If existing, this solution is independent of the initial condition and is thus frequently called a universal attractor. According to Barenblatt\cite{Barenblatt1996}, one of the remaining problems for which the application of this technique can yield novel and substantial results is that of the capillary-driven thin film equation that describes the dynamics of thin liquid films. In the past few years, many analytical\cite{Bowen2006,Myers1998,Ben13} and numerical studies\cite{Bertozzi1998,Salez2012a} have been performed in order to improve the mathematical knowledge of this equation and provide partial solutions to specific problems.

The dynamics of hole formation and growth has been extensively studied in the dewetting of thin viscous films\cite{Sro86, Bro90, Rei92, Rei93, Red94, Bro94, Xie98, See01b, Net04, Vil06, Bau09, Sno10, Bau12}. In the work by Stange \textit{et al.}~\cite{Sta97}, the growth of a small indentation in a 22 nm thick polystyrene (PS) film was followed. In this case, there is competition between the Laplace pressure\cite{deG03}, which acts to level any perturbation of the surface, and the disjoining pressure\cite{Her98, See01}, which amplifies surface modulations. When the disjoining pressure dominates, there is a continuous deepening of the hole and a subsequent rupturing and dewetting of the film. Alternatively, in a system where the Laplace pressure dominates over the disjoining pressure, capillary-driven levelling will take place. This latter phenomenon has recently been explored with a single step perturbing the surface of a PS film\cite{McG11, McG12, Sal12, Sal12b, McG13, Bom13}, where it was shown that the self-similar temporal evolution of the height profile can be used to accurately determine the capillary velocity (the ratio of the surface tension to the viscosity, $\gamma/\eta$) of the thin film, thus providing a simple and precise nanorheological probe. Furthermore, capillary levelling has been studied with a trench-like geometry\cite{Bau13, Ben13}, a patterned surface\cite{Lev08,Rog11,Tei11}, and a droplet spreading on an underlying viscous film\cite{Sal12, Cor12}. Crucial to these studies, and the one presented here, is that the length scales considered were well below the capillary length, thus the effect of gravity on the flow\cite{Hup82} can safely be neglected. 

Here, we report on the relaxation of nearly perfect cylindrical holes indented half-way into a thin PS film. This circular geometry is similar to that studied to probe dewetting, as described above, but in our case the films are thick enough that the samples are stable against spinodal dewetting, and disjoining forces can be ignored\cite{See01}. Instead, the capillary forces at the polymer-air interface dominate and the system will thus evolve towards a completely different final equilibrium steady state, namely that of a flat film instead of dispersed dewetted droplets. Films containing the cylindrical surface perturbations were annealed above the glass transition temperature, leading to a flow of the polymer melt and allowing for the capillary-driven relaxation to take place. By closely following the evolution of  several holes with different initial sizes, we find two distinct regimes in the approach to equilibrium: an early regime where opposing sides of the hole do not interact, and a late regime where the hole is filling up. In both cases, the energy dissipation of the system is well understood from quantitative scaling arguments and compared to that of previously studied geometries. Furthermore, profiles in the late regime converge perfectly to the analytically calculated universal asymptotic profile. Finally, while the evolution of single isolated holes is the core of this study, we have also investigated the relaxation of two holes that are in close proximity. Here, an interesting interplay between filling-up and coalescing of the holes was detected.

\section{Experiment}
\begin{figure}[t!]
\centering
\includegraphics[width=8cm]{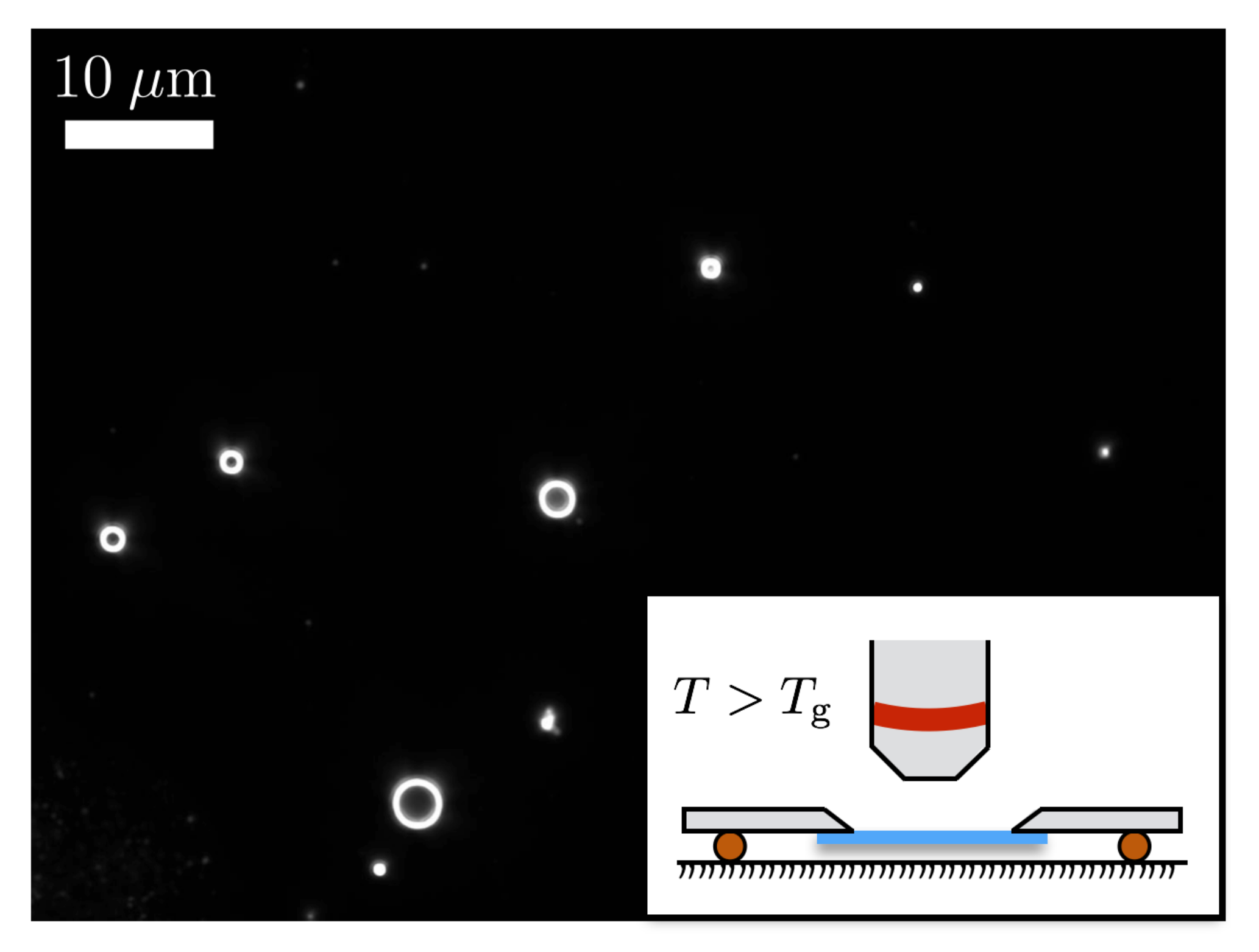}
\caption{Dark field optical microscopy image of a typical as-prepared sample, showing the rims of several randomly distributed circular holes with various sizes. The smallest holes are only visible as bright spots. The inset shows a schematic illustration (not to scale) of the experimental hole-formation setup. The washer, partially supporting the free-standing film, was placed on two rubber spacers on top of a heating stage and the film was then watched under a microscope as the holes formed and grew larger.}
\label{fig1}
\end{figure}

\begin{figure}[t!]
\centering
\includegraphics[width=5.5cm]{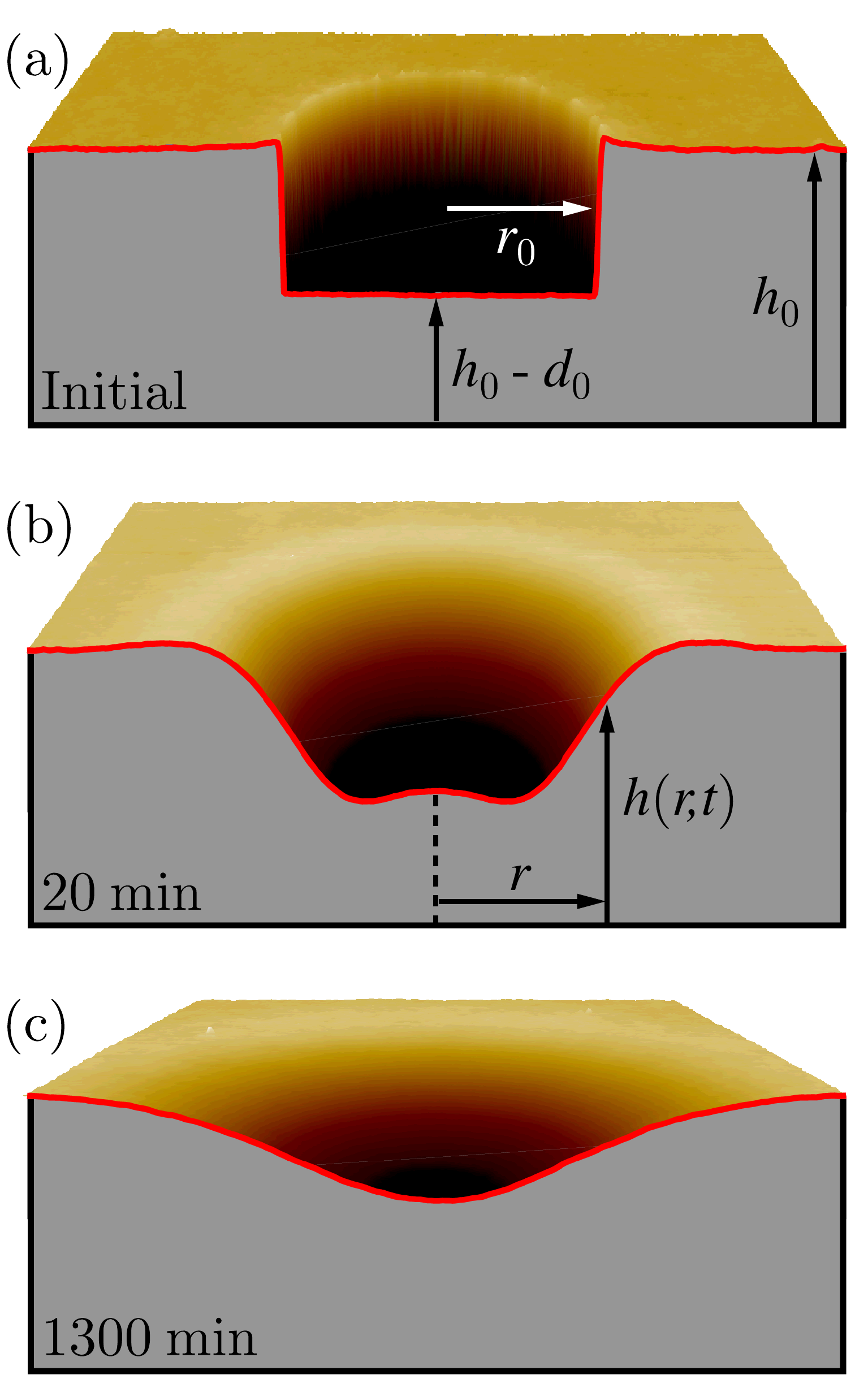}
\caption{(a) AFM surface topography of an initial hole with a radius of $r_0=2.8\ \mu$m and a depth of $d_0=75$~nm, in a film with a total thickness of $h_0=143$~nm. (b) and (c) show the same hole after 20 and 1300 min of total annealing at 140$^{\circ}$C. The horizontal size of all images is 15 $\mu$m and the thickness of the film in each radial point $r$, after a total annealing time $t$, is denoted by $h(r,t)$.}
\label{fig2}
\end{figure}

Polystyrene (PS, Polymer Source Inc.) films containing cylindrical holes, as shown in Figs.~\ref{fig1} and \ref{fig2}, were prepared in the following two stage process. First, PS films with a molecular weight of $M_{\textrm{W}} = 60\ \textrm{kg}.\textrm{mol}^{-1}$ and polydispersity index of $M_{\textrm{W}}/M_{\textrm{N}}=1.03$ were spin cast from a 2$\%$ toluene (Fisher Scientific, Optima grade) solution onto freshly cleaved mica sheets (Ted Pella Inc.)~as well as onto 1 cm $\times$ 1 cm silicon (Si) wafers (University Wafer). These films, prepared at the same time and from the same solution, all had the same nominal film thicknesses $h_{\textrm{mica}}=75\pm 1$~nm and $h_{\textrm{Si}}=67\pm 1$~nm, on the mica and Si substrates respectively. Previous to the spin casting, the Si wafers had been exposed to air plasma (Harrick Plasma, low power for 30 s) and rinsed in ultra pure water (18.2 M$\Omega$.cm, Pall, Cascada LS), methanol (Fisher Scientific, Optima grade), and toluene. All PS films were then pre-annealed in a vacuum oven ($\sim 10^{-5}$ mbar) for 24 hours at 130$^{\circ}$C, i.e. well above the glass transition temperature of PS ($T_{\textrm{g}}=97^{\circ}$C for the molecular weight used here), in order to remove residual solvent and relax the polymer chains.

In the second stage of the preparation, a pre-annealed film on the mica sheet was floated off the supporting sheet and onto the surface of ultra pure water and carefully picked up onto a 1 cm $\times$ 1 cm metal washer with a circular hole ($\diameter = 3$ mm) in the middle, thus creating a free-standing film with no supporting substrate. The washer was then placed, as shown in the inset of Fig.~\ref{fig1}, with the film facing downwards on a hot stage (Linkam) with two rubber spacers (thickness of a few mm) in between to reduce conductive heat transfer through the washer. The free-standing film was annealed  in air, 5-10$^{\circ}$C above the bulk $T_{\textrm{g}}$ and observed under a reflective microscope until small holes were formed through either nucleation on small dust particles or spinodal decomposition\cite{KDV00, Rot05}. Holes in free-standing films have been shown to grow exponentially in time\cite{KDV00,Deb95,Deb98}. Therefore, by waiting for a short while, holes with sizes up to tens of micrometers were produced, after which the washer was taken off the hot stage, quenched to room temperature, and placed (film down) atop the previously prepared sample with a PS film on Si substrate. Due to the strong van der Waals interaction between the two PS films, they contacted and the washer could then be carefully lifted off the sample, leaving the free-standing film (with cylindrical holes) atop the Si supported film. 

As shown in the dark field optical microscopy image of Fig.~\ref{fig1}, the as-prepared sample consists of a PS film with randomly distributed cylindrical holes of various sizes. As illustrated in Fig.~\ref{fig2}(a), the total film thickness and initial hole radius and depth are defined as $h_0$, $r_0$ and $d_0$, respectively. The ratio $d_0/(h_0-d_0)\approx 1.1$ was constant for all holes investigated, whereas the initial radius varied. 

To study the relaxation of these films, the samples were annealed for various amounts of time on a hot stage in air at 140$^{\circ}$C with a heating rate of 90$^{\circ}\textrm{C.min}^{-1}$. After each annealing cycle, the sample was taken off the hot stage and quenched to room temperature, thus inhibiting any further viscous flow (we note that the time to heat and quench samples is fast in comparison to the annealing times used). Holes were scanned with atomic force microscopy (AFM, Veeco Caliber) at room temperature both before any annealing had taken place (see initial AFM scan in Fig.~2(a)), as well as between every subsequent annealing cycle, as illustrated by the examples in Figs.~2(b) and 2(c). A large scale optical microscopy image (see Fig.~\ref{fig1}) of the sample was used as a map to find the same microscopic holes after each annealing cycle. The total thickness of the film at the radial position $r$, after a total annealing time $t$, is defined as $h(r,t)$. To determine the initial thickness of the bottom film ($h_0-d_0$), a sharp scalpel was used to make a scratch in the film down to the Si substrate, and the height of the step was then measured with AFM.

In experiments with long annealing times, profiles were difficult to access with AFM because of the shallow ($\le 20$ nm) and wide ($\sim 40 \ \mu$m) surface features. In such cases, imaging ellipsometry (Accurion Nanofilm EP3, Germany) was used. We ensured that both AFM and ellipsometry were in mutual agreement by carrying out calibration measurements with both tools. Radial height profiles were extracted from the centreline of the holes in order to investigate the temporal evolution of the relaxation caused by the capillary-driven viscous flow of the polymer melt.

\section{Theory}

Here we establish the theoretical framework in which the dynamics of the levelling holes can be understood. The evolution of the system is governed by a highly nonlinear equation that cannot be analytically solved. Nevertheless, by assuming particular self-similar symmetries and using conservation of a global quantity, the volume, we are able to characterise quantitatively the asymptotic behaviours of the system at short and long times. In particular, at long times we provide an exact analytical self-similar solution that is the universal attractor. Interestingly, here we explicitly take advantage of the intimate relation between scaling, self-similarity, and intermediate asymptotics \cite{Barenblatt1996}, to overcome the nonlinearity of the problem.

\subsection{Thin film equation}
Given that vertical length scales are small compared to typical radial scales, the use of the lubrication approximation is legitimate.
The levelling of the film is driven by gradients in the Laplace pressure and damped by viscosity. Given the height scales involved, hydrostatic\cite{Hup82} and disjoining pressures\cite{Her98, See01} can be safely neglected. The Stokes equation is used to connect local velocity and pressure. In addition, we assume radial symmetry at all times. Assuming no slip at the substrate and no stress at the free surface yields a Poiseuille flow in the radial direction. Finally, invoking incompressibility of the flow leads to the cylindrical capillary-driven thin film equation\cite{Sal12,Cor12}:
\begin{eqnarray}
\partial_t h +\frac{\gamma}{3\eta}\, \frac 1r \,\partial_r \left[   rh^3  \left(  \partial_r^{\,3} h +\frac 1r\,  \partial_r^{\,2} h  -\frac{1}{r^2}\, \partial_r h  \right) \right]&=&0 \ .\label{CTFE}
\end{eqnarray}
Nondimensionalization of lengths is performed as follows:
$h = Hh_0, \, d_0 = D_0 h_0,\,  r = R r_0$. Consistently with Eq.~\eqref{CTFE}, time is nondimensionalized as: $t =   {\,  3 T \eta r_0^4}/{(\gamma h_0^{\,3})}$. This leads to the dimensionless cylindrical capillary-driven thin film equation:
\begin{eqnarray}
\,\,\,\,\partial_T H + \frac 1R \,\partial_R \left[   RH^3  \left(  \partial_R^{\,3} H +\frac 1R\,  \partial_R^{\,2} H  -\frac{1}{R^2}\, \partial_R H  \right) \right]\,\,\,\,=\,\,\,\,0 \ .\label{CTFEad}
\end{eqnarray}
An interesting quantity to describe the evolution of the profile is the excess surface $S$, with respect to the one of the flat equilibrium configuration, as this quantity is proportional to the excess surface energy  of the film $\gamma S$. We define the dimensionless excess surface $\tilde {S}={S}/(\pi h_0^2)$, which for small slopes is well approximated by:
\begin{eqnarray}
\tilde {S} &=& \int_0^\infty \text d R\,R\, \left( \partial_RH\right)^2  \ . 
\label{Surface}
\end{eqnarray}

In the following Sections, we focus on understanding the short-term and long-term asymptotic behaviours of physical quantities such as the just defined excess surface, the perimeter of the hole, and the full height profile itself. Note that although all the functions here also depend parametrically on the vertical initial aspect ratio $D_0$, we do not write explicitly this dependence for the sake of clarity, and because the reported experiments are performed with almost a unique $D_0$.

\subsection{Early time evolution}
\label{ete}
At early times, we expect the central depth of the profile to be constant and the edge front of the hole not to be influenced by the central region ($R\sim0$). Therefore, guided by previous studies\cite{McG12,Sal12,Sal12b,Bau13}, we centre the short-term profiles at the edge of the hole through the change of variables: $H(\check R+1,T)=\check H(\check R,T)$. Together with Eq.~\eqref{Surface}, this leads to:
\begin{eqnarray}
\tilde {S}
&=&\int_{-1}^{\infty} \text d \check R\, (\check R+1) (\partial_{\check R} \check H)^2 \ . \label{Surf1}
\end{eqnarray}
Consistently with the dimensional analysis of Eq.~\eqref{CTFEad}, and with the short-term result of a previous study on linear trenches\cite{Bau13}, we assume short-term self-similarity of the profile in the variable $\check U=\check RT^{-1/4}$, \textit{with constant amplitude}. This translates into: 
\begin{equation}
\label{stss}
\check H(\check U\cdot T^{1/4},T)=\check F(\check U)\ ,
\end{equation}
where $\check F$ is an unknown function. Performing such a change of variables in Eq.~\eqref{Surf1} leads to:
\begin{eqnarray}
\tilde {S}&=&\int_{-T^{-1/4}}^{\infty} \text d \check U \left(\check U+\frac1{T^{1/4}}\right)\check F'( \check U)^2 \ . \label{Surf2}
\end{eqnarray}
Using $T\ll 1$, one obtains the further simplification of Eq.~\eqref{Surf2}:
\begin{eqnarray}
\tilde {S}&\approx& \frac1{T^{1/4}}\int_{-\infty}^{\infty} \text d \check U\,\check F'( \check U)^2 \ , 
\label{Surf3}
\end{eqnarray}
since the nonzero information of $\check F'$ is located around the edge front ($\check U\sim0$). This predicts a short-term temporal scaling of the excess surface, and thus excess energy, in $T^{-1/4}$.

\subsection{Late time evolution}
\label{lte}
In the long-term regime, we now expect the central depth of the profile to reduce in time until the flat equilibrium situation is reached. Therefore, we set $H(R,T)=1+\Delta(R,T)$, where the new relevant unknown $\Delta(R,T)$ is the excess profile with respect to the final reference state. It shall be treated as a perturbation later on. Before, consistently with the dimensional analysis of Eq.~\eqref{CTFEad}, and guided by the long-term results of previous studies\cite{Bau13,Ben13}, we only assume long-term self-similarity of the profile in the variable $U=RT^{-1/4}$, \textit{with a decaying amplitude}. This translates into:
\begin{eqnarray}
\Delta(R,T)&=&\alpha(T)G(U)\ , 
\label{prod}
\end{eqnarray}
where $G$ and $\alpha$ are, as yet, unknown functions, and where we impose $G(0)=1$ with same generality. Note that this form is different from the short-term self-similarity of Eq.~(\ref{stss}) as we now allow for an amplitude factor $\alpha(T)$ that enables the filling-up of the hole. In order to determine the function $\alpha(T)$, we consider the dimensionless volume $\tilde {V}$ of the perturbation. Consistent with the nondimensionalization of lengths detailed in part 2.1, the dimensionless volume is defined from the real volume $V$ of the perturbation as $\tilde {V}={V}/(2\pi r_0^2h_0)$, and therefore satisfies: 
\begin{eqnarray}
\tilde {V} &=& \int_0^\infty \text d R\,R\, \Delta (R,T)\ . 
\label{Volume}
\end{eqnarray}
Performing the change of variables of Eq.~(\ref{prod}) into Eq.~\eqref{Volume} leads to: 
\begin{eqnarray}
\tilde {V} &=& \alpha(T) \,T^{1/2}\int_0^\infty \text d U\,U\, G (U)  \ .
\end{eqnarray}
Then, since the volume has to be conserved, one has: 
\begin{eqnarray}\alpha(T)=\beta\,T^{-1/2}\ ,\label{G}\end{eqnarray}
where $\beta=\alpha(1)=T^{1/2}\Delta(0,T)$ is a constant factor. 

According to Eqs.~\eqref{Surface}, \eqref{prod}, and \eqref{G}, one obtains:
\begin{eqnarray}
\tilde {S} &=&\frac{\beta^2}{T} \int_0^\infty \text d U\,U\, G' (U)^2  \ ,
\label{Surflate}
\end{eqnarray}
thus predicting a long-term temporal scaling for the excess surface, and thus excess energy, in $T^{-1}$. 

Another relevant geometrical quantity is the dimensionless perimeter (or radius, equivalently) $P_{\theta}$ of the profile where $\theta$ is a given height ratio with respect to the central depth of the hole. It can be defined from the real perimeter $p_{\theta}$, through $P_{\theta}=p_{\theta}/(2\pi r_0)$, and from the relation:
\begin{equation}
\label{tet}
\Delta(P_\theta,T)=\theta \Delta(0,T)\ .
\end{equation}
Invoking Eqs.~\eqref{prod} and \eqref{tet}, one can write: $G\left(P_{\theta}\,T^{-1/4}\right)=\theta$. Inverting the function $G$ is allowed when $\theta$ is chosen so that $P_{\theta}$ is uniquely defined. In this case, one has:
\begin{eqnarray}
P_\theta&=& T^{1/4}\,G^{-1}(\theta) \ .
\label{perimeq}
\end{eqnarray}
Since the factor $G^{-1}(\theta)$ depends only on the fixed threshold $\theta$, Eq.~(\ref{perimeq}) predicts a long-term temporal scaling of the perimeter in $T^{1/4}$. 

Finally, we show that in the long-term regime one can also determine analytically the function $G$. Indeed, at late times the hole is filling up and, at some point, the surface displacement can reasonably be considered as a small perturbation of the film: $\Delta(R,T)=H(R,T)-1\ll 1$. In this regime, it is thus legitimate to linearise Eq.~\eqref{CTFEad}:
\begin{eqnarray}
\left[ \partial_T +\partial_R^{\,4} +\frac2R\,\partial_R^{\,3}  -\frac{1}{R^2}\,\partial_R^{\,2}  +\frac1{R^3} \partial_R \right] \Delta&=&0\ . \label{CTFElin}
\end{eqnarray}
Combining Eqs.~\eqref{prod}, \eqref{G}, and \eqref{CTFElin}, leads to the \textit{ordinary} differential equation:
\begin{eqnarray}
G''''+\frac {2G'''} U-\frac{G''}{U^2}+\left(  \frac 1{U^3}   -\frac U4\right)G' -\frac G2 &=&0 \ .\label{CTFEU}
\end{eqnarray}
Equation~\eqref{CTFEU} can be solved analytically together with the natural boundary conditions of the problem: $\lim_{U\rightarrow \infty} G(U) =\lim_{U\rightarrow \infty} G'(U)=G'(0)=0$ and $G(0)=1$. Its solution reads:
\begin{eqnarray}
G(U)=\frac{U^2}{4\sqrt \pi}\ _0H_{2}\left[\left\{ \frac32,\frac32\right\},\frac{U^4}{256}\right]-\ _0H_{2}\left[\left\{ \frac12,1\right\},\frac{U^4}{256}\right] ,
\label{HPG}
\end{eqnarray}
where the $(0,2)$-hypergeometric function\cite{Abramowitz1965,Gradshteyn1965} is defined as: 
\begin{eqnarray}
_0H_{2}\left[\left\{ a,b\right\},w\right]&=&\sum_{k\geq0} \frac{1}{(a)_k(b)_k}\,\frac{w^k}{k!}\ ,  
\end{eqnarray}
with the Pochhammer notation $(.)_k$ for the rising factorial. This long-term self-similar solution is expected to be independent of the initial condition\cite{Barenblatt1996}. It is thus called the universal attractor, or the intermediate asymptotic solution, of Eq.~\eqref{CTFEad}. Note that this universal solution is strongly linked to the spatial dimension and boundary conditions of the problem. For a detailed derivation together with experimental evidence of the 2D long-term solution in regard of different boundary conditions, see previous studies\cite{Ben13,Bau13,McG12,Sal12b}.

\section{Results and Discussion}

We now turn to the results of the experiments described above and compare to the theoretical predictions. We first focus on single individual holes, before studying the coalescence of binary systems.

\subsection{Single holes}

For clarity, we first focus on three single holes of different initial radii: a small one ($r_0=0.8$ $\mu$m), a medium one ($r_0=2.8$ $\mu$m), and a large one ($r_0=4.4$ $\mu$m). We study the overall behaviour with the medium hole, and use the large and small holes to address the short-term and long-term regimes respectively. 

\begin{figure}[t!]
\centering
\includegraphics[width=8cm]{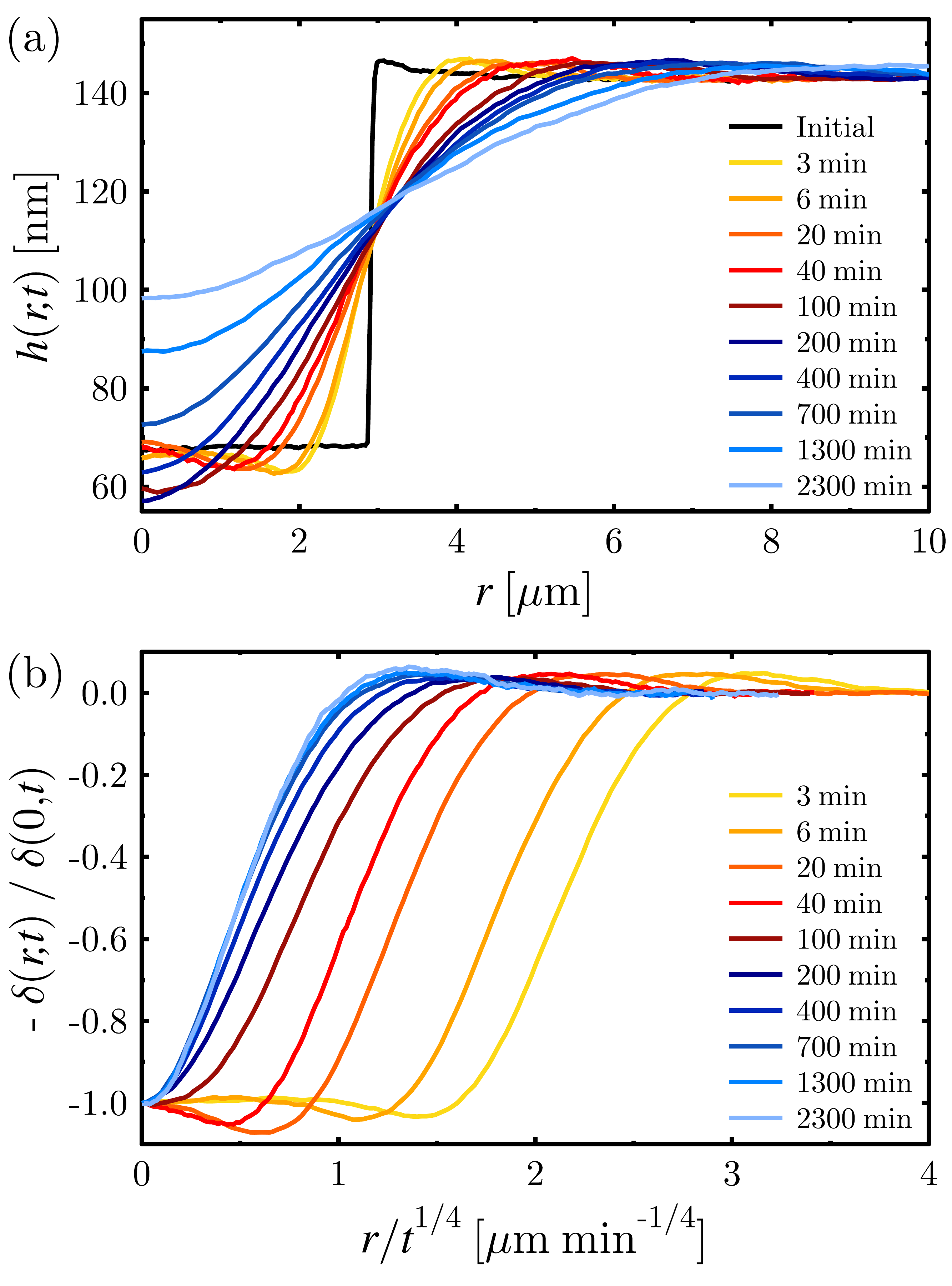}
\caption{(a) Height profiles of the medium hole ($r_0=2.8$ $\mu$m) as a function of radial distance $r$, after different total annealing times $t$. (b) Normalised height profile, $\delta(r,t)=h(r,t)-h_0$, of the same hole, as a function of the long-term self-similar variable.}
\label{fig3}
\end{figure}

In Fig.~\ref{fig3}(a), the height profile of the medium hole (also introduced in Fig.~\ref{fig2}) is plotted as a function of radial distance, for several total annealing times. The ``cuspy'' rim of the initial hole is the result of the hole growth mechanism in viscoelastic films close to $T_{\textrm{g}}$\cite{Rei01, Sau02, She02}. This cusp is flattened out on time scales significantly shorter than those relevant to our experiment, due to the large Laplace pressure associated with such a geometry. Similarly, at early times, the corners of the hole relax by forming a ``dip'' at the bottom and a ``bump'' at the top, thus evening out quickly these sharp features of the cylindrical structure. The profile broadens with time similar to that seen with single steps\cite{McG11,Sal12b}, trenches\cite{Bau13}, and droplets\cite{Cor12}. As time progresses, material flows in from the sides of the film and fills up the hole in a symmetric fashion. In Fig.~\ref{fig3}(b), the normalised height for the same hole is plotted as a function of the long-term self-similar variable $u=r\,t^{-1/4}$ predicted in the Theory section. It can be seen how the normalised dip of the wave deepens before the long time relaxation stage occurs. In the latter regime, the collapse of the profiles at late times becomes apparent (see profiles at 700, 1300, and 2300 min in Fig.~\ref{fig3}(b)). 

\begin{figure}[t!]
\centering
\includegraphics[width=8cm]{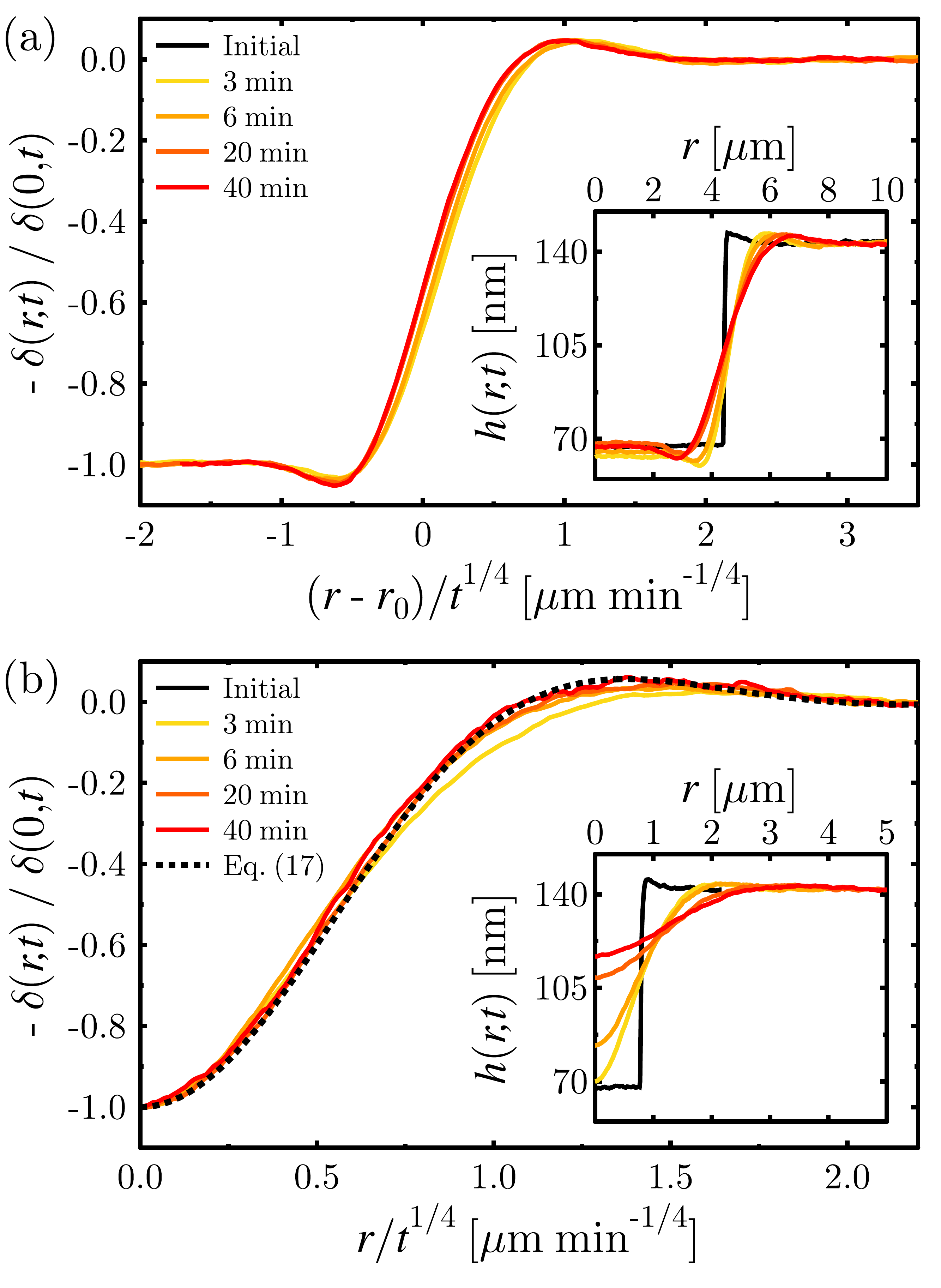}
\caption{(a) Short-term normalised height profiles, $\delta(r,t)=h(r,t)-h_0$, for the large hole ($r_0=4.4$ $\mu$m), as a function of the short-term self-similar variable. (b) Long-term normalised height profiles for the small hole ($r_0=0.8$ $\mu$m), as a function of the long-term self-similar variable. The analytical attractor of Eq.~\eqref{HPG} has been fit to the collapsed profiles. In both (a) and (b), the insets show the corresponding non-normalised profiles.}
\label{fig4}
\end{figure}

A larger initial hole radius makes it more precise to study the early time that lasts longer. In Fig.~\ref{fig4}(a), the short-term normalised height profiles of the large hole are plotted as a function of the short-term self-similar variable $(r-r_0)\,t^{-1/4}$ of the theory. They collapse onto one single profile, indicating self-similarity in the early time regime consistently with the theoretical assumptions of Section~\ref{ete}.

For a smaller initial hole radius, the late time regime is reached faster and this system thus allows for the study of the long-term asymptotic regime. In Fig.~\ref{fig4}(b), the long-term normalised height profiles of the small hole are plotted as a function of the long-term self-similar variable $r\,t^{-1/4}$ of the theory. They collapse onto one single profile that is well fitted by the analytical attractor of Eq.~\eqref{HPG}, thus validating the theoretical assumptions and predictions of Section~\ref{lte}. Moreover, this single parameter fit (horizontal stretch) allows for a measurement of the inverse capillary velocity of the film: $\eta/\gamma=0.27\ \textrm{min}.\mu\textrm{m}^{-1}$, which is consistent with previously measured values for $M_\textrm{W}=60\ \textrm{kg}.\textrm{mol}^{-1}$ PS thin films\cite{McG13}.

\begin{figure}[t!]
\centering
\includegraphics[width=8cm]{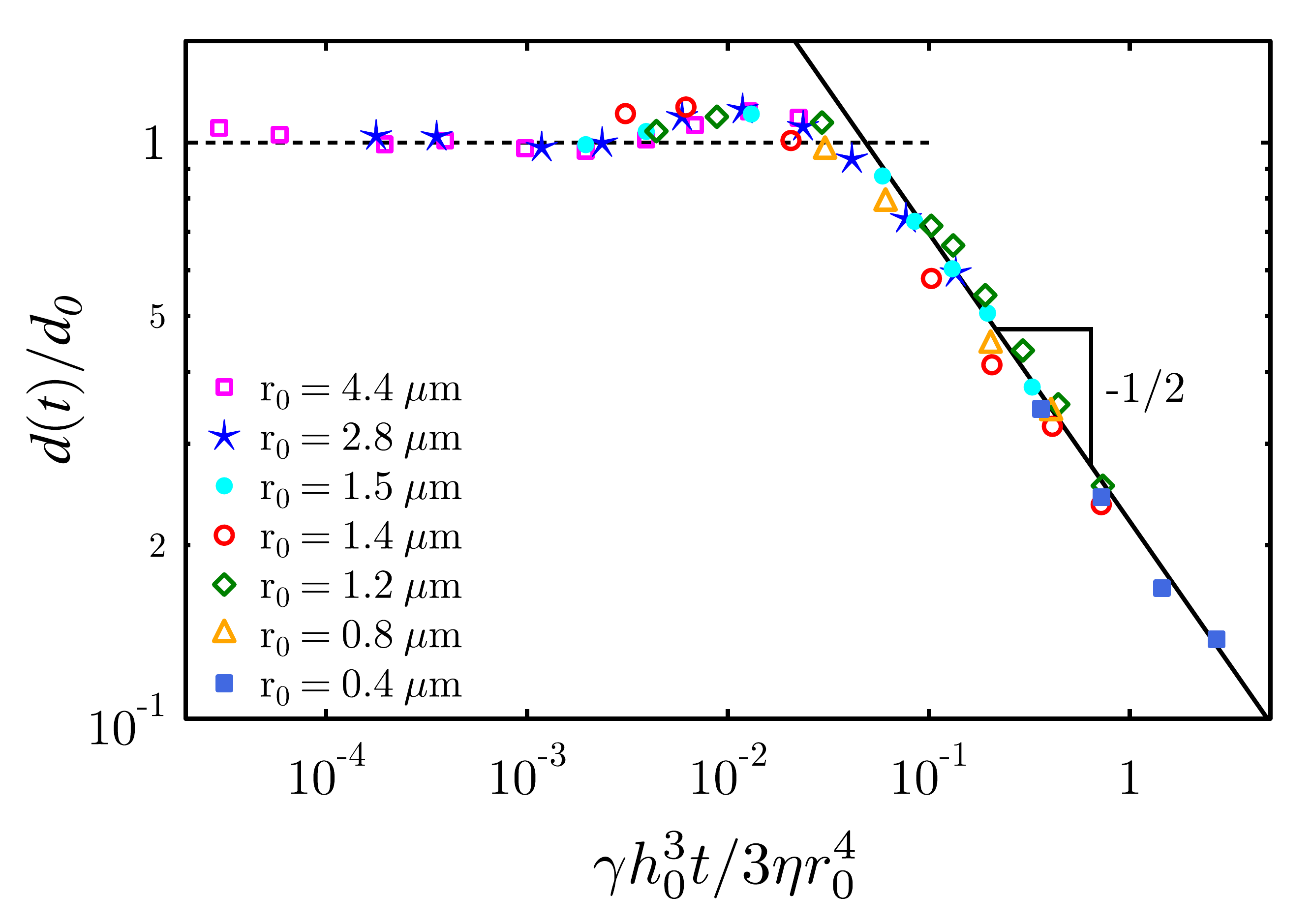}
\caption{Normalised depth as a function of normalised time for all holes studied in this work (markers show different holes with initial radii $r_0$). The dashed line shows the initial normalised depth whereas the solid line is the theoretical prediction for the scaling in the late time regime, as given by Eqs.~\eqref{prod} and \eqref{G}.}
\label{fig5}
\end{figure}

We now consider all the holes that have been experimentally studied, including the three previously described. As indicated by Eqs.~\eqref{stss}, \eqref{prod}, and \eqref{G}, the hole depth $d(t)=h_0-h(0,t)$ is expected to be constant at small times and to scale like $T^{-1/2}$ at late times, where $T= \gamma h_0^{\,3}t/(\,  3 \eta r_0^4)$. Figure~\ref{fig5} shows the normalised depth as a function of $T$ for all holes studied in this work, resulting in a collapse of all data sets, as expected from the theory. The crossover in the temporal scaling demonstrates the strong influence of the boundary conditions on the levelling dynamics. In the early time regime, the sides of the hole are not interacting and the depth remains unchanged. As the opposite sides of the hole start to interact, the hole centre counterintuitively moves far below the initial depth of the hole, which causes a small but measurable ($\log$-$\log$) shoulder in the crossover regime of Fig.~\ref{fig5}. At long times, the holes fill and evolve eventually towards the equilibrium state corresponding to a completely flat film with no gradients in the Laplace pressure and thus no more capillary-driven flow. As indicated by the dashed and solid lines, the experimental data confirms the constant-depth assumption of the short-term theory and is in excellent agreement with the predicted long-term temporal scaling in $T^{-1/2}$.

\begin{figure}[t!]
\centering
\includegraphics[width=8cm]{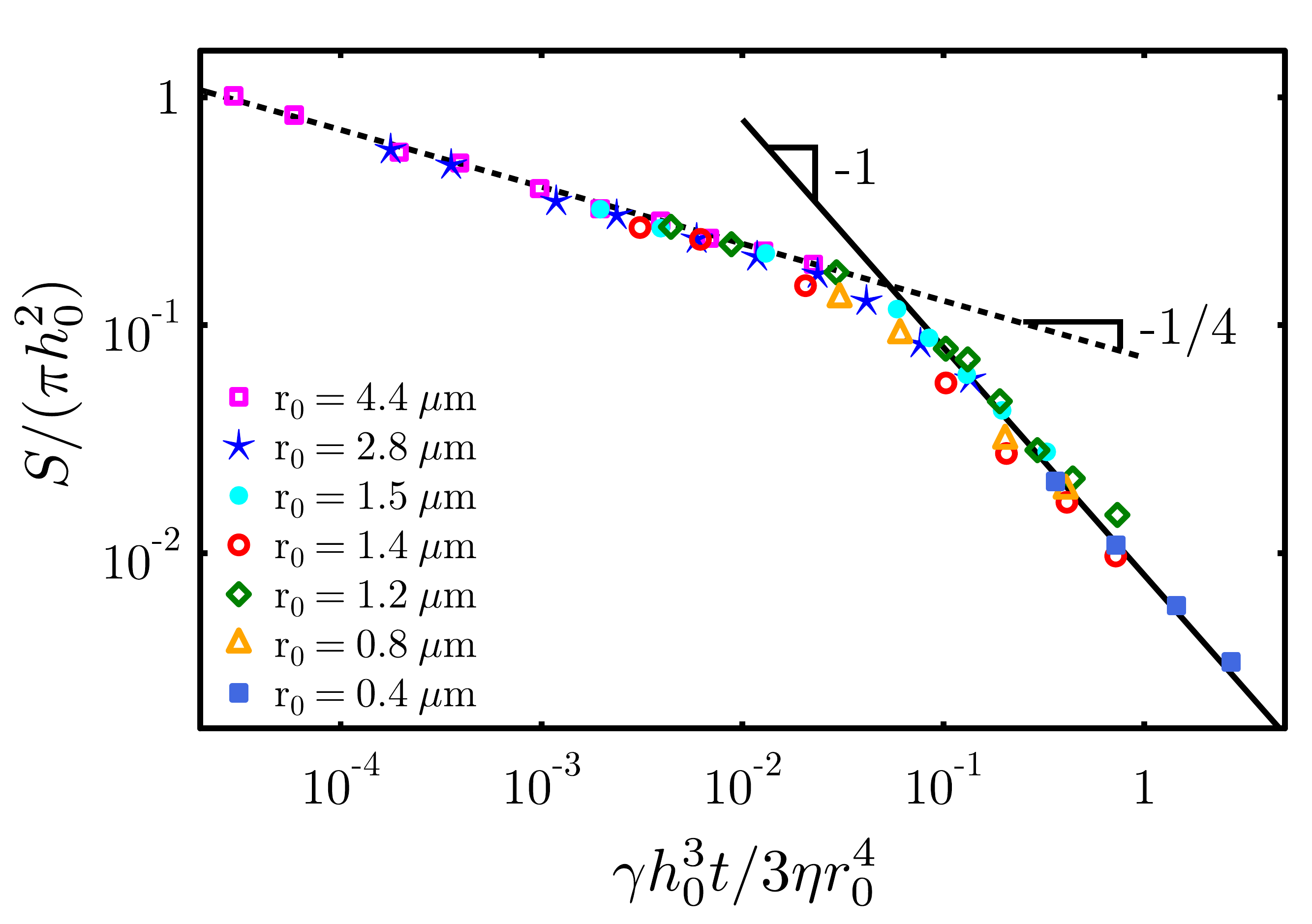}
\caption{Normalised excess surface area as a function of normalised time. The lines show the early (dashed line) and late (solid line) time regime scalings predicted by Eqs.~\eqref{Surf3} and \eqref{Surflate}.}
\label{fig6}
\end{figure}

The relevant quantity to study the global excess capillary energy of the system is the excess surface area $S$, as defined in Eq.~\eqref{Surface}. It was calculated for all holes, between all annealing steps, and is plotted in Fig.~\ref{fig6} as a function of normalised time. Again, the data collapses in full agreement with the theoretical quantitative scaling laws for the early and late temporal regimes, as formulated in Eqs.~\eqref{Surf3} and \eqref{Surflate} respectively. Apart from the change in dissipation rate between both regimes, linked to the change in effective boundary conditions, there is a continuous decrease of the total surface area, and thus global surface energy, which directly results from the total bulk dissipated viscous power. The somewhat counterintuitive local flow features seen at the bottom of the holes in the crossover regime, as pointed out in Fig.~\ref{fig5}, are simply the result of the local constructive interference of opposing sides of the holes and do not affect the global decrease of energy reported in Fig.~\ref{fig6}.

The early $T^{-1/4}$ scaling is identical to that observed in the levelling of a single 2D step\cite{McG12,Sal12b} and in the short-term relaxation of a 2D trench\cite{Bau13}. This can be understood as follows: at early times the edge front of the hole is far form the centre which brings the problem close to a 2D geometry, i.e. that of linear trenches and steps. However, the late time energy dissipation of the 3D holes reported here is significantly faster than that of the 2D trenches at long times: $T_{\textrm{hole}}^{-1}$ vs. $T_{\textrm{trench}}^{-3/4}$. This fundamental difference is due to the three-dimensionality of the hole, for which material is allowed to flow in from all azimuthal orientations, resulting in a faster relaxation process. The transition between the two regimes also occurs earlier for the holes than what was observed with the 2D trenches\cite{Bau13}. 

\subsection{Double holes}
\begin{figure}[t!]
\centering
\includegraphics[width=8cm]{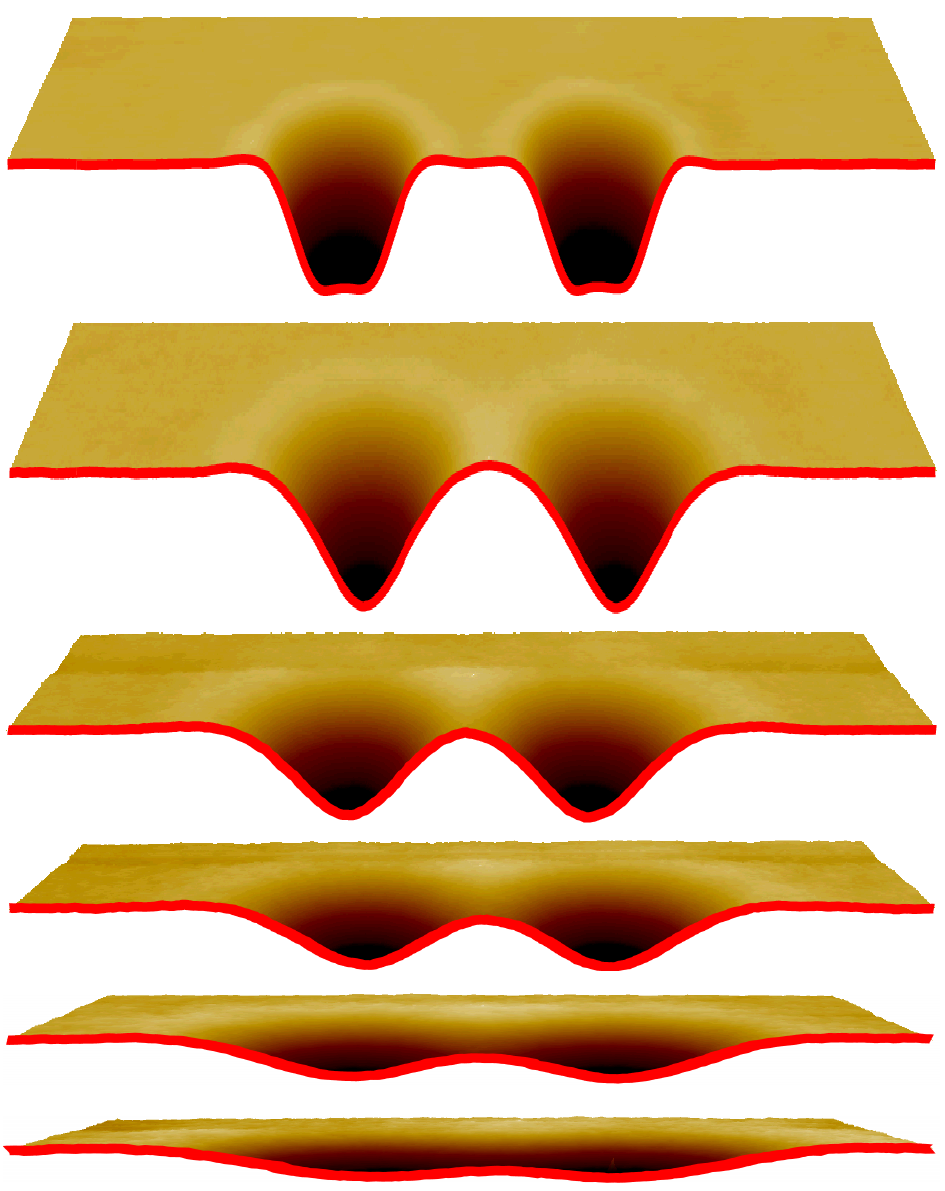}
\caption{AFM surface topographies of two neighbouring identical holes after 2, 30, 120, 240, 480, and 900 minutes of total annealing at 140$^{\circ}$C. The horizontal size of all images is 25 $\mu$m and the vertical scale is the same for all images.}
\label{fig7}
\end{figure}
\begin{figure}[t!]
\centering
\includegraphics[width=8cm]{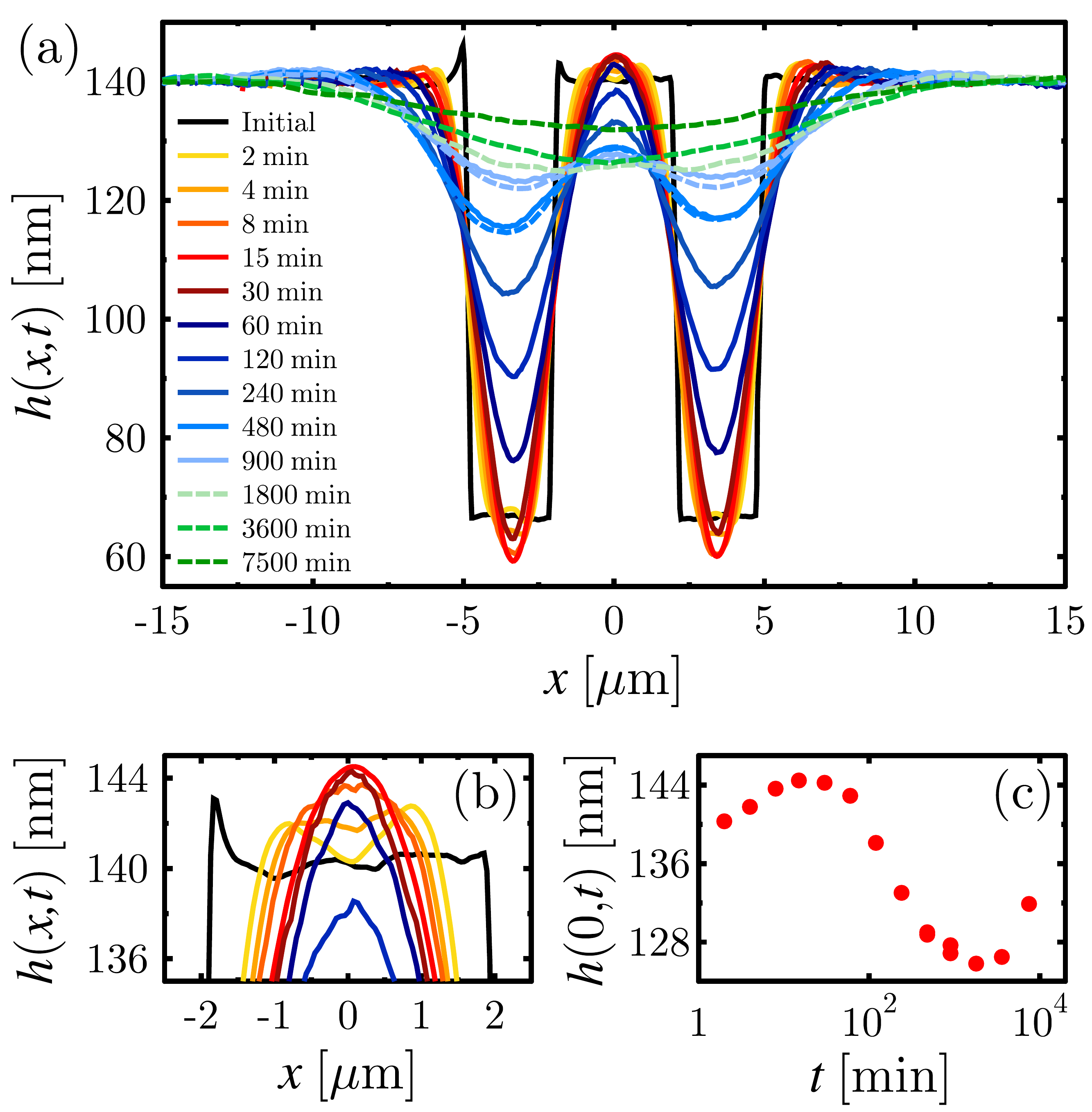}
\caption{(a) Temporal evolution of the height profile of the two neighbouring (edges initially separated by $4\ \mu$m) holes with an initial average radius and depth of $r_0=1.4$ $\mu$m and $d_0= 74$ nm, respectively. The dashed lines indicate measurements performed with the imaging ellipsometer. (b) Short-term profiles in the bridge region between the holes. (c) Vertical position of the middle of the bridge as a function of time.}
\label{fig8}
\end{figure}

In this work, we also studied a binary system consisting, coincidentally, of two nearly equally sized holes located a small distance apart, as shown in the temporal surface topography series of Fig.~\ref{fig7}. The two neighbouring holes have edges initially separated by $4\ \mu$m and initial average radius and depth of $r_0=1.4$ $\mu$m and $d_0= 74$ nm, respectively. The height profile of this pair is plotted in Fig.~\ref{fig8}(a) as a function of the lateral distance $x$ and for several annealing times $t$. At early times, the two holes can be seen to evolve independently of each other. As the holes widen out, they start to interact and coalesce. Figure~\ref{fig8}(b) shows the features of the initial profiles in the inter-hole region (``bridge''), and material can be seen to be pushed in and up by both of the widening holes. As shown in Fig.~\ref{fig8}(c), when the holes come close enough the bridge between the two moves down, causing material to flow out. This continues until the bridge reaches the same level as the rising centres of the individual holes. Afterwards, the coalesced hole is slowly filling up again while changing from an elliptical to an axisymmetric shape. When the latter symmetry is recovered, the overall dynamics is expected to be described by Eq.~\eqref{CTFE}. 

\begin{figure}[t!]
\centering
\includegraphics[width=8cm]{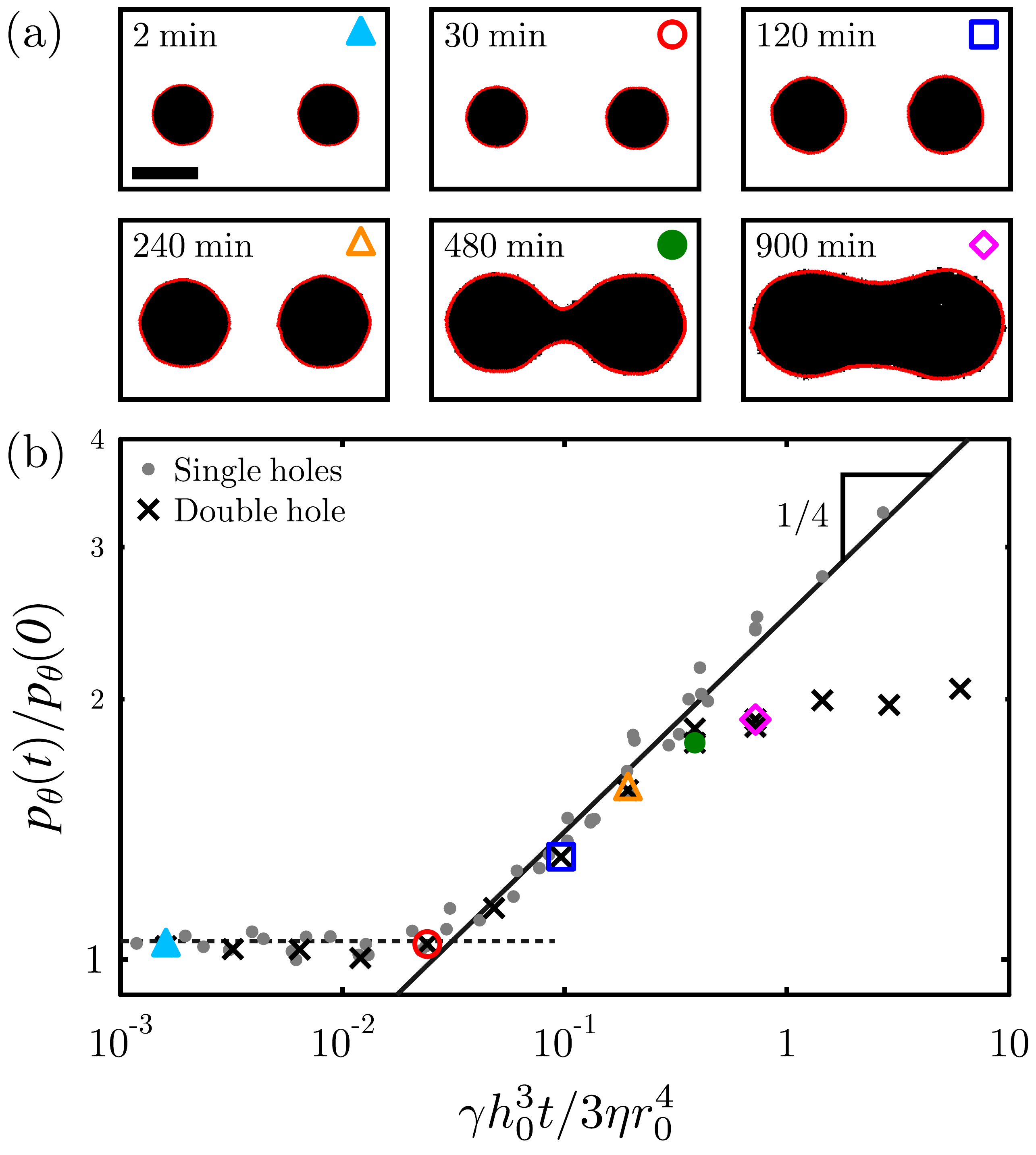}
\caption{(a) Top view images of the double hole system after the same annealing times as in Fig.~\ref{fig7}, with the perimeter drawn in red. The scale bar is 3 $\mu$m. (b) Normalised perimeter as a function of normalised time. Both single (all holes studied) and double hole AFM and ellipsometry data are included. The six individual (colour) markers correspond to the double hole data of (a).}
\label{fig9}
\end{figure}

The coalescence can be seen in the AFM surface plots of Fig.~\ref{fig7}, as well as in the top view images of Fig.~\ref{fig9}(a). To quantify this process in more detail, the temporal evolution of the perimeter $p_\theta$ of the binary system was calculated from experimental data and compared to that of single holes. The perimeter was defined to be vertically positioned at $\theta = 40\ \%$ of the central depth (see Section~\ref{lte}). This reference fraction corresponds to the approximate intersection point for all non-normalised short-term profiles (see e.g.  Fig.~\ref{fig3}(a)). The results are shown in Fig.~\ref{fig9}(b), where the normalised perimeter is plotted as a function of the normalised time. 

The single hole data collapse on a master curve, and a transition between the initial and final regimes is observed at the same normalised time as for the depth and excess surface area in Figs.~\ref{fig5} and \ref{fig6}, as expected by the theory. In the early time regime, no perimeter change occurs which is simply the result of the particular choice of the reference fraction $\theta$. As the holes transition to the long-term regime, the perimeter grows with a $T^{1/4}$ power law as predicted by Eq.~\eqref{perimeq}. Naturally, this long-term scaling law is now independent of the choice of the reference fraction $\theta$, since it corresponds to the overall self-similar behaviour described in Section~\ref{lte}.

As far as the perimeter of the double hole system is concerned, it follows initially the single hole scaling perfectly. As shown in Fig.~\ref{fig9}, the hole perimeter remains unchanged up to 30 minutes of total annealing, after which each individual hole enters the \textit{single hole} long-term regime and thereby widen out according to the $T^{1/4}$ prediction of Eq.~\eqref{perimeq}. Afterwards, a strong deviation from the single hole behaviour can be seen when the two holes start to interact and coalesce (full green circle). The temporal location of this departure from the single hole behaviour depends obviously on the definition of the reference fraction $\theta$, but always occurs when the coalescence becomes apparent in a top view similar to the one of Fig.~\ref{fig9}(a). The initial distance between the holes must also influence the temporal location of this departure. For holes of identical size, and for a fixed $\theta$, one expects the departure location to simply depend on the ratio of the hole radius to the hole separation. By increasing this ratio, the departure should occur earlier in time and could in principle even happen before the single hole crossover to the single-hole long-term $T^{1/4}$ regime. Finally, at very large times, we expect the coalesced binary hole to become axisymmetric again. Then, the overall dynamics is expected to be described again by Eq.~\eqref{CTFE}, and thus the departure branch of Fig.~\ref{fig9} to join back the single-hole long-term $T^{1/4}$ regime.

\section*{Conclusions}
We have reported on the capillary-driven relaxation of cylindrical holes in thin polystyrene films above their glass transition temperature. Two different self-similar regimes were predicted through quantitative scaling arguments, and were successfully confirmed experimentally. In addition, we calculated analytically the universal self-similar attractor of the developed thin film theory and showed the excellent agreement with the long-term experimental profiles. At a given initial depth of the hole, the time scales involved in the two relaxation regimes were shown to depend only on the initial radius of the hole, the thickness of the film, and the capillary velocity of the material. The excess energy of the system was calculated and shown to be monotonically decreasing as the system approached the equilibrium configuration. Moreover, at the crossover between the two self-similar regimes, the temporal excess energy scaling was shown to change from $t^{-1/4}$ to $t^{-1}$, reflecting the crucial influence of the effective boundary conditions and of the dimension of the system. Finally, a double hole binary system was investigated and a clear deviation from the single hole behaviour was observed during coalescence. 

The excellent agreement between theory and experiments demonstrates the efficiency of the intermediate asymptotics theory, in a new cylindrical 3D geometry, to overcome the nonlinearity of the governing thin film model and to provide quantitative scalings for important global quantities in the system, such as the total energy. The understanding of the evolution of this simple system has implications for polymer-based industrial applications where the existence of small topological features or defects have the ability to enhance or destroy the function of associated devices. Finally, the reported system provides an ideal probe of the relaxation and coalescence processes that act in thin viscous films. 

\section*{Acknowledgements}
The financial supports by NSERC of Canada, the Swedish Cultural Foundation in Finland, and \'Ecole Normale SupŽ\'erieure of Paris, are gratefully acknowledged. The authors also thank Joshua McGraw and Oliver B\"{a}umchen for interesting discussions.
\balance
\footnotesize{
\providecommand*{\mcitethebibliography}{\thebibliography}
\csname @ifundefined\endcsname{endmcitethebibliography}
{\let\endmcitethebibliography\endthebibliography}{}

}

\end{document}